# On Estimating the First Frequency Moment of Data Streams

Sumit Ganguly
IIT Kanpur
India

Purushottam Kar
IIT Kanpur
India

October 30, 2018

## Abstract

Estimating the first moment of a data stream defined as $F_1 = \sum_{i \in \{1,2,\ldots,n\}} |f_i|$ to within $1 \pm \epsilon$-relative error with high probability is a basic and influential problem in data stream processing. A tight *space* bound of $O(\epsilon^{-2} \log(mM))$ is known from the work of [9]. However, all known algorithms for this problem require per-update stream processing time of $\Omega(\epsilon^{-2})$, with the only exception being the algorithm of [6] that requires per-update processing time of $O(\log^2(mM)(\log n))$ albeit with sub-optimal space $O(\epsilon^{-3} \log^2(mM))$. [1]

In this paper, we present an algorithm for estimating $F_1$ that achieves near-optimality in both space and update processing time. The space requirement is $O(\epsilon^{-2}(\log n + (\log \epsilon^{-1}) \log(mM)))$ and the per-update processing time is $O((\log n) \log(\epsilon^{-1}))$.

## 1 Introduction

The data stream model serves as an abstraction for a variety of monitoring applications, including, data networks, sensor networks, financial data, etc.. In this model, an input stream $\sigma$ is abstracted as a potentially infinite sequence of records of the form $(pos, i, v)$, where, $i \in \{1, 2, \ldots, n\} = [n]$ and $v \in \mathbb{Z}$ is the change to the frequency $f_i$ of item $i$. The *pos* attribute is simply the sequence number of the record. Each input record $(pos, i, v)$ changes $f_i \leftarrow f_i + v$. Thus, $f_i = \sum_{(\text{pos},i,v)} v$, that is, $f_i$ is the sum of the changes made to the frequency of $i$ since the inception of the stream. The vector $f = [f_1, f_2, \ldots, f_n]^T$ is called the frequency vector of the stream.

The $p$th frequency moment is defined as $F_p = \sum_{i \in [n]} |f_i|^p$. The problem of estimating $F_p$, and in particular, the estimation of $F_0$, $F_1$ and $F_2$, have been fundamental to the development of data stream processing techniques and lower bounds. In this paper, we consider the problem of estimating $F_1$ to within approximation factor of $1 \pm \epsilon$ and with probability at least some constant $c > 0.5$, where the probability is taken over the internal random bits used by the algorithm. We will say that a randomized algorithm computes an $\epsilon$-approximation to a real valued quantity $L$, provided, it returns $\hat{L}$ such that $|\hat{L} - L| < \epsilon L$, with probability that is at least some absolute constant strictly larger than $1/2$. Since prior work [1] shows that any deterministic algorithm for 0.1-approximation of $F_p$, $p \geq 0$ requires $\Omega(n)$ space, we consider the problem of randomized $\epsilon$-approximation of $F_1$.

---

[1] At the IIT Kanpur Workshop on Algorithms for Massive Data Sets, Dec 18-20 2009, Jelani Nelson announced the discovery of an algorithm (with David Woodruff) for estimating $F_1$ that uses space $O(\epsilon^{-2} \log^{O(1)}(mM))$ space and time $O(\log^{O(1)}(mM))$. Since their work is unpublished, we are unable to make a comparison.

We assume that items come from the domain $[n] = \{1, 2, \ldots, n\}$, each stream update $(pos, i, v)$ has $|v| \leq M$ and the size of the stream is $m$ i.e. the number of records appearing in the stream. [1] presents a seminal randomized sketch technique for $\epsilon$-approximation of $F_2$ in the data streaming model using space $O(\epsilon^{-2} \log(mM))$ bits. Estimation of $F_0$ (i.e., the number of $i \in [n]$ s.t. $|f_i| \neq 0$) was first considered by Flajolet and Martin in [4] and improved in [1, 7, 2]. Since the techniques for estimating $F_p$ for $p > 2$ are substantially different from those used for estimating $F_p$ for $0 < p \leq 2$, we do not review this line of work.

## 1.1 Review: Previous work on estimating small moments

We now review existing work on estimating $F_p$, for $p \in (0, 2]$. In terms of lower bounds for estimating $F_p$, Woodruff [13] presents an $\Omega(\epsilon^{-2})$ space lower bound for the problem of estimating $F_p$, for all $p \geq 0$. This is improved to $\Omega(\epsilon^{-2} \log(\epsilon^2 M))$ in [9].

The notation $X \sim D$ means that the random variable $X$ has probability distribution $D$. The term *i.i.d.* stands for independent and identically distributed family of random variables.

**Indyk's estimator.** The use of $p$-stable sketches was pioneered by Indyk [8] for estimating $F_p$ for $0 < p \leq 2$. A $p$-stable sketch is a linear combination $X = \sum_{i=1}^{n} a_i s_i$ where the $s_i$'s are drawn independently from the $p$-stable distribution $St(p, 1)$ with scale factor 1. By property of stable distributions, $X \sim St\left(p, (F_p(a))^{1/p}\right)$. For estimating $F_1$, keep $t = O(\frac{1}{\epsilon^2})$ independent 1-stable (i.e., Cauchy) sketches $X_1, X_2, \ldots, X_t$ and let $\hat{F}_1 = (4/\pi) \cdot \text{median}_{r=1}^{t} |X_r|^q$. Then, $\hat{F}_1 \in (1 \pm \epsilon) F_1$ with probability $15/16$. Further, Indyk shows that for stable distributions it suffices to, (a) truncate the support of the distribution $St(p, 1)$ beyond $(mM)^{O(1)}$, and, (b) consider the approximation to the continuous $St(p, 1)$ distribution by discretizing it using into a grid with interval size $(mM/\epsilon)^{-O(1)}$.

To reduce the number of random bits required to maintain independent sketches, Nisan's pseudo-random generator (PRG) [11] is used for fooling space $S$ bounded randomized machine computation–here $S = O(\epsilon^{-2} \log(\epsilon^{-1} mM))$. We can assume that the stream is ordered since the sketches are linear and therefore their values are independent of the order of item arrivals. For each element $i$, the stable random variables $s_i(u)$ for $u = 1, 2, \ldots, t$ are computed from the $i$th chunk of $S$ random bits obtained from Nisan's generator that stretches a seed of length $S \log n$ to $nS$ bits. The time taken to obtain the $i$th random bit chunk is $O(\epsilon^{-2} \log(\epsilon^{-1})(\log n))$ simple field operations on a field of size $O(mM\epsilon^{-1})$. Kane, Nelson and Woodruff [9] observe that a seed length of $O(\log(\frac{mM}{\epsilon}) \log(n))$ suffices.

**Li's estimator.** Li [10] proposes several new estimators for the estimation of $F_p$ for $p \in (0, 2)$, most notably the geometric means estimator. These estimators are defined on $p$-stable sketches $X_u = \sum_{i \in [n]} f_i s_i(u)$, $u = 1, 2, \ldots, t$. The geometric means estimator is defined as

$$\hat{Y}_{p,t} = C(p, p/t)^{-t} \prod_{i=1}^{t} |X_i|^{p/t}.$$

where

$$C(p, q) = \frac{2}{\pi} \Gamma\left(1 - \frac{q}{p}\right) \Gamma(q) \sin\left(\frac{\pi q}{2}\right), \qquad -1 < q < p\ .$$

Li [10] proves that $(i)$ the estimator is unbiased, that is, $\mathsf{E}\left[\hat{Y}_{p,t}\right] = F_p$, and, $(ii)$ $|\hat{Y}_{p,t} - F_p| < \epsilon F_p$ with probability $1/8$ provided, $t = \Omega(\epsilon^{-2})$.

**Other work.**  Kane, Nelson and Woodruff [9] present algorithms for estimating $F_p$ for $p \in (0, 2)$ that use space that is tight with respect to the lower bounds. The update processing time is is $O(\epsilon^{-2}(\log \epsilon^{-1})^2/(\log \log \epsilon^{-1}))$ simple operations on fields of size $(mM)^{O(1)}$.

An estimator for $F_p$ based on the HSS technique was presented in [6] for estimating $F_p$. Though it uses sub-optimal space $O(\epsilon^{-2-p}(\log(mM)^2(\log n)))$, it has the best update processing time so far, namely, $O(\log^2(mM))$.

## 1.2 Contributions

We present a novel algorithm for estimating $F_1$ that is nearly optimal with respect to both space and update-processing time. So far, all known algorithms, except the HSS based technique [6] have a per-update processing time of $\Omega(\epsilon^{-2})$. The HSS technique however is sub-optimal in space and requires space $O(\epsilon^{-3}(\log(mM))^2(\log n))$ for estimating $F_1$. In this paper, we present an algorithm for estimating $F_1$ whose resource usage is nearly optimal in terms of *both* space and time. The space requirement of our algorithm is $O\bigl((\epsilon^{-2}(\log(n\epsilon^{-1})))\log(mM) + (\log n)(\log \epsilon^{-1})\log(mM)\bigr)$. The time for processing each stream update is $O\bigl((\log n)(\log \epsilon^{-1})\bigr)$ simple operations on $O(\log(mM))$ bit numbers. [2]

## 2 Algorithm for estimating $F_1$

In this section, we present an algorithm for estimating $F_p$ that has fast update time. We first describe the data structure and then the estimator.

*Notation.* $F_p^{\text{res}}(k)$ is defined as follows. Let $|f_{s_1}| \geq |f_{s_2}| \geq \ldots \geq |f_{s_n}|$. Then $F_p^{\text{res}}(k) = \sum_{j=k+1}^{n}|f_{s_j}|^p$.

Let $\varepsilon$ be the user-supplied accuracy parameter and set $\epsilon = \varepsilon/10$.

STABLESKETCH **and** COUNTSKETCH **structure.**  The STABLESKETCH structure is a hash table $U$ having $C = 64B$ buckets numbered from 1 to $64B$, where, $B = 1/\epsilon^2$ and having a hash function $h : [n] \to [C]$ that is chosen uniformly at random from a hash family $\mathcal{H}$ mapping $[n] \to [C]$. The degree of independence required of the hash family will be determined later; for now, it is assumed to be fully independent.

For $b \in [C]$ each bucket $U[b]$ of the tables maintains three linear $p$-stable sketches denoted by $X_{b,1}$, $X_{b,2}$ and $X_{b,3}$ as follows.

$$X_{b,r} = \sum_{i=1}^{n} f_i s_{b,r}(i), \qquad b \in [C], r \in \{1, 2, 3\} \ .$$

For each value of $b$ and $r$, the random variables $\{s_{b,r}(i)\}_{i \in [n]}$ are independent (this independence will be relaxed later). For each value of $b$, the seeds for the random variables $s_{b,r}(i)$ and $s_{b,r'}(i')$, for $r \neq r'$ are three-wise independent. Across buckets in the same table, the stable sketches need only to be pair-wise independent, that is the seeds for the random variables $s_{b,r}(i)$ and $s_{b',r'}(i')$, for $b \neq b'$ are pair-wise independent. The sketches are updated corresponding to each update $(i, v)$ as follows.

$$X_{j,h(i),r} := X_{j,h(i),r} + v \cdot s_{j,b,r}(i), \quad r = 1, 2, 3 \ .$$

We keep a COUNTSKETCH structure [3] consisting of $g$ hash tables $T_1, T_2, \ldots, T_g$, where $g = O(\log \frac{1}{\epsilon^2})$ and each table consists of $C$ buckets. Later, the degree of independence is determined and reduced. Heavy hitters are identified using (another) COUNTSKETCH structure, denoted as $\text{HH}_2^C$, that can return an estimate $\hat{f}_i$ of the frequency $f_i$ such that $|\hat{f}_i - f_i| \leq 8\left(\frac{F_2^{\text{res}}(C/8)}{C}\right)^{1/2}$, with constant probability of success $127/128$. We let this COUNTSKETCH structure to have $O(\log n)$ independent hash tables and functions. The COUNTSKETCH data structures together use a total space of $O(\epsilon^{-2}(\log n + \log(\epsilon^{-1})))$ bits. The time taken to update this structure is $O(\log n + \log \epsilon^{-1})$.

---
[2]See footnote on Page 1

## 2.1 Estimator

*Estimating $F_2^{\text{res}}$.* The algorithm of [5] is applied to the $\text{HH}_2^C$ data structure to obtain estimates for $F_2^{\text{res}}(\epsilon B)$ and $F_2^{\text{res}}(B)$ that are accurate to factors of $1 \pm 1/128$ with prob. at least $127/128$.

**Heavy and light items.** After estimating $F_2^{\text{res}}(B)$, we estimate the frequencies of all heavy-hitters. Items are classified according to their estimated frequencies into two categories as follows.

$$(i) \text{ heavy: } \hat{f}_i^2 \geq \frac{4\hat{F}_2^{\text{res}}(4B)}{B} \quad \text{and} \quad (ii) \text{ light: } \hat{f}_i^2 < \frac{4\hat{F}_2^{\text{res}}(4B)}{B} \ . \tag{1}$$

The set of heavy and light items are denoted respectively as $H$ and $L$. The algorithm obtains separate estimates for the contribution to $F_p$ from the heavy items and the light items, and adds them to obtain the final estimate. That is,

$$\hat{F}_p = \hat{F}_p^H + \hat{F}_p^L \ .$$

*Notation.* For any set $R \subset [n]$, let $F_p(R)$ denote $\sum_{i \in R}|f_i|^p$.

The true contributions of the items in $H$ and $L$ are as follows: $F_p^H = F_p(H), \quad F_p^L = F_p(L)$ .

**Heavy estimator.** We identify the set $H$ of heavy items as those elements whose estimated frequencies satisfy (1)(i). Say that the event NOHVYCOLL($i$) holds if there is some table index $j \in [g]$ such that no other heavy item maps to the same bucket as $h_j(i)$. That is,

$$\text{NOHVYCOLL}(i) \equiv \exists j \in [g] \text{ s.t. } \forall k \in H \setminus \{i\}, h_j(i) \neq h_j(k), \ .$$

If NOHVYCOLL($i$) holds, then, let $\theta(i)$ denote the index $j \in [g]$ such that $i$ is isolated from all other heavy items in its bucket for table $T_j$.

For $i \in H$ we obtain an estimate as follows. If NOHVYCOLL($i$) holds, then, $\theta(i)$ exists and let $b = h_{\theta(i)}(i)$ be the bucket to which $i$ maps to under $h_{\theta(i)}$. Also, let $\xi_j$ be the AMS 4-wise independent hash function mapping items to $\{1, -1\}$ corresponding to table $T_j$. The estimate is obtained as

$$Y_i = \begin{cases} T_j[b] \cdot \text{sgn}(\hat{f}_i) \cdot \xi_j(i) & \text{if NOHVYCOLL}(i) \text{ holds, where, } j = \theta(i), b = h_j(i) \\ 0 & \text{otherwise.} \end{cases} \tag{2}$$

The heavy estimate is: $\hat{F}_1^H = \sum_{i \in H} Y_i$.

**Light Estimator.** For bucket index $b \in [C]$ say that the event NOCOLLSION($b$) holds if no heavy item maps to bucket $b$ in table $U$. That is

$$\text{NOCOLLSION}(b) \equiv \forall k \in H, h(k) \neq b \ .$$

The estimate returned is

$$\hat{F}_p^L = C_L \sum_{b \in \mathcal{B}} (C(p, p/3))^{-3} |X_{b,1}|^{p/3} |X_{j,b,2}|^{p/3} |X_{j,b,3}|^{p/3}$$

where, $C_L = 1/\Pr[\textsc{NoCollsion}(b)] = (1 - 1/C)^{-|H|}$.

The final estimator is the sum of heavy and light estimators, namely, $\hat{F}_1 = \hat{F}_1^H + \hat{F}_1^L$.

## 3 Analysis

Throughout this section, we will assume that $\epsilon \leq 1/8$, $B = \epsilon^{-2}$ and $C = 64B$.

**Claim 1** $|H| \leq 5.1B$ *with probability* $127/128$.

**Proof** See Appendix A.

The following lemma is standard from arguments in tail bounds of frequency powers.

**Lemma 3.1** *Suppose* $|f_{s_1}| \geq |f_{s_2}| \geq \ldots \geq |f_{s_n}|$. *Then, for any* $0 < p \leq q$,

$$\sum_{j=B+1}^{n} |f_{s_i}|^q \leq \frac{1}{B^{q/p-1}} \left( \sum_{j=1}^{n} |f_{s_i}|^p \right)^{q/p} . \tag{3}$$

*In particular, for* $q = 2p$, $\sum_{j=B+1}^{n} |f_{s_i}|^{2p} \leq \frac{1}{B} \left( \sum_{j=1}^{n} |f_{s_i}|^p \right)^2$.

**Proof** See Appendix A.

### 3.1 Analysis of Light Estimator

The light estimator $\hat{F}_p^L$ is analyzed in the general setting when $p \in (0, 2)$.

Let $\mathcal{B}$ be the set of buckets in table $U$ such that no element of $H$ maps to any of these buckets, that is, $\mathcal{B} = \{b \in [C] \mid \forall i \in H, h_j(i) \neq b\}$.

**Lemma 3.2** $\mathsf{E}\left[\hat{F}_p^L\right] = F_p^L$.

**Proof**

$$\mathsf{E}_{h,s}\left[\hat{F}_p^L\right] = C_L \mathsf{E}_h \left[ \sum_{b \in \mathcal{B}} \sum_{h(i)=b} |f_i|^p \mid h \right] = C_L \sum_{i \notin H} |f_i|^p \cdot \Pr[h_j(i) \in \mathcal{B}] = \sum_{i \in L} |f_i|^p \quad \blacksquare$$

Define

$$K_p = (C(p, p/3))^{-6} (C(p, 2p/3))^3 \text{ where, } C(p, q) = \frac{2}{\pi} \Gamma\left(1 - \frac{q}{p}\right) \Gamma(q) \sin\left(\frac{\pi q}{2}\right) .$$

As shown by Li [10], $K_p \leq (\pi^2/36)(p^2 + 2) + 1 \leq 2.5$.

Random variables such as $F_p^L$ are functions of two independent sets of random bits, namely, the hash function $h$ and the bits used by the stable variables denoted as $s$. To explicitly denote this dependence, we will denote by notations such as $\mathsf{Var}_{h,s}\left[F_p^L\right]$ and $\mathsf{E}_{h,s}(F_p^L)$ the variance and expectation of $F_p^L$ (or any suitable random variable) over the random seeds of $h$ and $s$. Then notation $\mathsf{E}_s[F_p^L]$ is used to emphasize that the expectation is taken over the random bits of $s$, by holding the random bits of $h$ fixed. In effect this is the same as $\mathsf{E}[F_p^L \mid h]$. Therefore, $\mathsf{E}\left[F_p^L\right] = \mathsf{E}_h\left[\mathsf{E}_s\left[F_p^L\right]\right]$, since the random bits used by $h$ and $s$ are independent.

**Lemma 3.3** $\text{Var}_{h,s}\left[F_p^L\right] \leq (K_p C_L - 1)\sum_{i\in L}|f_i|^{2p} + \frac{K_p C_L}{C}\left(\sum_{i\in L}|f_i|^p\right)^2$ .

**Proof of Lemma 3.3** Denote the estimate of bucket $b \in \mathcal{B}$ obtained from the light estimator to be $Y_b = C_L\big(C(p,p/3)\big)^{-3}|X_{b,1}|^{p/3}|X_{b,2}|^{p/3}|X_{b,3}|^{p/3}$. Then,

$$Y = \hat{F}_p^L = \sum_{b\in\mathcal{B}} Y_b = \sum_{b\in\mathcal{B}} C_L\big(C(p,p/3)\big)^{-3}|X_{b,1}|^{p/3}|X_{b,2}|^{p/3}|X_{b,3}|^{p/3} . \tag{4}$$

Let $C_L$ be the probability that an item $i \in L$ does not conflict with any item in $H$ under the hash function $h_j$. Under full independence of $h$, $C_L = (1-1/C)^{|H|}$.

We have,

$$\mathsf{E}_{h,s}\left[Y^2\right] = \sum_{b\in\mathcal{B}}\mathsf{E}_{h,s}\left[Y_b^2\right] + \sum_{b,b'\in\mathcal{B}, b\neq b'}\mathsf{E}_{h,s}\left[Y_b Y_b'\right] . \tag{5}$$

Let $b \in \mathcal{B}$ and $K_p = (C(p,p/3))^{-6}(C(p,2p/3))^3$.

$$\mathsf{E}_h\left[\mathsf{E}_s\left[\sum_{b\in\mathcal{B}(h)} Y_b^2 \,\bigg|\, h\right]\right] = K_p C_L^2 \mathsf{E}_h\left[\sum_{b\in\mathcal{B}(h)}\left(\sum_{h(i)=b}|f_i|^p\right)^2 \bigg| h\right] = K_p C_L^2 \mathsf{E}_h\left[\sum_{h(i)=b}\left(|f_i|^{2p} + \sum_{\substack{i\neq i' \\ h(i)=h(i')=b}}|f_i f_{i'}|^p\right)\bigg| h\right]$$

$$= K_p C_L^2 \sum_{i\in L}|f_i|^{2p} \cdot \Pr[\text{NoCollsion}(i)]$$

$$+ K_p C_L^2 \sum_{i\neq i', i,i'\in L}|f_i f_{i'}|^p \cdot \Pr\left[h(i) = h(i'), \text{NoCollsion}(i)\right]$$

$$= K_p C_L \sum_{i\in L}|f_i|^{2p} + \frac{K_p C_L}{C}\left(\sum_{i\in L}|f_i|^p\right)^2 \tag{6}$$

Further, for $b \neq b'$, and $b,b' \in \mathcal{B}$,

$$\mathsf{E}_{h,s}\left[\sum_{b\neq b'} Y_b Y_{b'}\right] = \mathsf{E}_h\left[\mathsf{E}_s\left[\sum_{b\neq b'} Y_b Y_{b'}|h\right]\right] = \mathsf{E}_h\left[\mathsf{E}_s[Y_b|h]\mathsf{E}_s[Y_{b'}|h]\right], \text{ since, } b\neq b' \text{ and full indep. of } h.$$

$$= C_L^2 \sum_{i\neq i'}|f_i|^p|f_{i'}|^p \Pr\left[h(i) \neq h(i'), \text{NoCollsion}(i), \text{NoCollsion}(j)\right]$$

$$\leq \left(\sum_{i\in L}|f_i|^p\right)^2 - \sum_{i\in L}|f_i|^{2p} \tag{7}$$

since $\Pr[h(i) \neq h(i'), \text{NoCollsion}(i), \text{NoCollsion}(j)] = (1-1/C)(1-2/C)^H \leq (1-1/C)C_L^2$.

Substituting (6) and (7) into (5), we get

$$\text{Var}_{h,s}[Y] = \mathsf{E}_{h,s}\left[Y^2\right] - \left(\sum_{i\in L}|f_i|^p\right)^2 \leq (K_p C_L - 1)\sum_{i\in L}|f_i|^{2p} + \frac{K_p C_L}{C}\left(\sum_{i\in L}|f_i|^p\right)^2 . \blacksquare$$

**Lemma 3.4** $\left|\hat{F}_p^L - F_p^L\right| \leq 6(1.75/8^{1-p/2} + 5/16)^{1/2}\epsilon F_p$ with prob. $35/36$ .

**Proof**

$$\sum_{i\in L}|f_i|^{2p} \leq \left(\max_{i\in L}|f_i|\right)^p \sum_{i\in L}|f_i|^p \leq \left(\frac{F_2^{\text{res}}(8B)}{B}\right)^{p/2} F_p \leq \frac{1}{B^{p/2}(8B)^{1-p/2}}F_p^2 = \frac{\epsilon^2 F_p^2}{8^{1-p/2}} \tag{8}$$

since, $B = 1/\epsilon^2$. Further, $K_p \leq (\pi^2/36)(p^2 + 2) + 1 \leq 2.5$ and $C_L \leq (1 - |H|/C)^{-1} \leq (1 - 5.1B/64B)^{-1} \leq 1.1$ by Claim 1. Therefore, by Lemma 3.3 and (8), we have

$$\mathsf{Var}\big[\hat{F}_p^L\big] \leq (K_p C_L - 1) \sum_{i \in L} |f_i|^{2p} + \frac{K_p C_L}{C} \Big(\sum_{i \in L} |f_i|^p\Big)^2 \leq \big(1.75/8^{1-p/2} + 2.75/64\big) \epsilon^2 F_p^2$$

By Chebychev's inequality,

$$\Pr\Big[\big|\hat{F}_p^L - F_p^L\big| > 6\big(1.75/8^{1-p/2} + 2.75/64\big)^{1/2} \epsilon F_p\Big] \leq \frac{1}{36} \;. \quad \blacksquare$$

### 3.2 Analysis of Heavy Estimator

In this section, we analyze the heavy estimator for estimating $F_1^H$.
For any set $K \subset [n]$, let $F_2^{\text{res}}(K) = F_2 - F_2(K) = \sum_{i \notin K} |f_i|^2$. The following lemma is from [5].

**Lemma 3.5** *Let $K$ be the items that are top-k with respect to estimated absolute frequencies using the* COUNTSKETCH *algorithm with table height $64B$. Let $|K| = k$ and suppose* TOP-K($k$) *be the indices of the top-k items of $f$ w.r.t. absolute frequencies. If $k \leq 8B$, then, $F_2^{\text{res}}(k) \leq F_2^{\text{res}}(K) \leq F_2^{\text{res}}(k)\big(1 + 2\sqrt{k} + k\big)$.*

**Proof of Lemma 3.5.**

$$F_2^{\text{res}}(K) = \sum_{i \notin K} |f_i|^2 = \sum_{i \notin (\text{TOP-K}(k) \cup K)} |f_i|^2 + \sum_{i \in \text{TOP-K}(k), i \notin K} f_i^2$$

$$= \sum_{i \notin (\text{TOP-K}(k) \cup K)} |f_i|^2 + \sum_{i \in \text{TOP-K}(k) \setminus K} f_i^2$$

$$\leq \sum_{i \notin (\text{TOP-K}(k) \cup K)} |f_i|^2 + \sum_{i \in K \setminus \text{TOP-K}(k)} (f_i + \Delta)^2$$

$$\leq \sum_{i \notin (\text{TOP-K}(k) \cup K)} |f_i|^2 + \sum_{i \in K \setminus \text{TOP-K}(k)} f_i^2 + 2\Delta \sum_{i \in K \setminus \text{TOP-K}(k)} |f_i| + |K \setminus \text{TOP-K}(k)|\Delta^2$$

$$= F_2^{\text{res}}(k) + 2\Delta |K \setminus \text{TOP-K}(k)|^{1/2} \Big(\sum_{i \in K \setminus \text{TOP-K}(k)} |f_i|^2\Big)^{1/2} + \frac{|K \setminus \text{TOP-K}(k)| F_2^{\text{res}}(8B)}{B}$$

$$\leq F_2^{\text{res}}(k) + 2 \frac{(|K \setminus \text{TOP-K}(k)| F_2^{\text{res}}(8B))^{1/2}}{B} \big(F_2^{\text{res}}(k)\big)^{1/2} + k F_2^{\text{res}}(8B)$$

$$\leq F_2^{\text{res}}(k) + 2\sqrt{k} F_2^{\text{res}}(k) + k F_2^{\text{res}}(k) \quad \blacksquare$$

For a heavy item $i \in H$, let NOHVYCOLL($i$) be the event that $i$ does not collide with any of the other heavy items in one of the buckets in the COUNTSKETCH structure tables $T_1, \ldots, T_g$, that is,

$$\text{NOHVYCOLL}(i) \equiv \exists r \in [g] \text{ s.t. } \forall k \in H \setminus \{i\}, h_r(k) \neq h_r(i)$$

The event NOHVYCOLL($H$) is said to occur if NOHVYCOLL($i$) occurs for each $i \in H$. That is,

$$\text{NOHVYCOLL}(H) \equiv \forall i \in H, \text{NOHVYCOLL}(i) \text{ holds.}$$

Assuming full independence, $\Pr[\text{NOHVYCOLL}(H)] \geq 1 - |H|\big(1 - \big(1 - \frac{1}{C}\big)^{|H|-1}\big)^g$. Since, $|H| \leq 5.1B$, $C = 64B$, we have $\Pr[\text{NOHVYCOLL}(i)] \geq \frac{31}{32}$ if $g \geq \frac{\log 32|H|}{\log(2(|H|-1)/C)}$. Since $|H| \leq 5.1B$, it suffices to let $g = 2 + \log \frac{5.1}{\epsilon^2}$.

If NOHVYCOLL($i$) holds then let $j = \theta(i)$ be the index of (some) $r \in [g]$ such that $i$ has no collision with any item of $H$ (except itself) under the hash function $h_j$. Let $T = T_{\theta(i)}$, $h = h_{\theta(i)}$ and $\xi = \xi_{\theta(i)}$. Then, let

$$Y_i = C_H \cdot T_{\theta(i)}[h_{\theta(i)}(i)] \cdot \text{sgn}(f_i) \cdot \xi_{\theta(i)}(i) \ .$$

Although we do not know $\text{sgn}(f_i)$ we can use $\text{sgn}(\hat{f}_i)$ instead which is equal to it with very high probability.

**Lemma 3.6** *For $i \in H$, $\mathsf{E}[Y_i] = |f_i|$. Thus, $\mathsf{E}\left[\sum_{i \in H} Y_i\right] = F_1^H$.*

**Proof**

$$\mathsf{E}_\xi[Y_i \mid \text{NOHVYCOLL}(H)] = \mathsf{E}\Big[f_i \cdot \text{sgn}(i) \cdot \xi(i)^2 + \sum_{h(k)=h(i), k \neq i} f_k \xi(j)\xi(i)\text{sgn}(i)\Big]$$
$$= f_i \cdot \text{sgn}(i) = |f_i| \ . \quad \blacksquare$$

**Lemma 3.7** *Let $i, k \in H$, $i \neq k$. Then, $\mathsf{E}[Y_i Y_k \mid \text{NOHVYCOLL}(H)] = |f_i||f_k|$.*

**Proof** Let $i \neq j$ and consider $Y_i Y_j$. Assume that NOHVYCOLL($H$) holds. Then,

$$Y_i Y_j = \big(T_{\theta(i)}[h_{\theta(i)}(i)] \cdot \text{sgn}(f_i) \cdot \xi_{\theta(i)}(i)\big) \cdot \big(T_{\theta(j)}[h_{\theta(j)}(j)] \cdot \text{sgn}(f_j) \cdot \xi_{\theta(j)}(j)\big) \ .$$

There are two cases, namely, either (i) $\theta(i) = \theta(j)$ or (ii) $\theta(i) \neq \theta(j)$.

If $t = \theta(i) \neq \theta(j) = t'$, then,

$$Y_i Y_j = \big(\text{sgn}(f_i) \sum_{i':h_t(i)=h_{t'}(i')} f_{i'} \xi_t(i)\xi_t(i')\big) \cdot \big(\text{sgn}(f_j) \sum_{j':h_{t'}(j')=h_{t'}(j)} f_{j'} \xi_{t'}(j)\xi_{t'}(j')\big)$$

Since $t \neq t'$, the two multiplicands use independent random bits, since $\{\xi_t\}$ are independent of $\{\xi_{t'}\}$'s. Hence, the expectation of the product is the product of the expectations, the conditional on NOHVYCOLL($H$) notwithstanding. Therefore,

$$\mathsf{E}[Y_i Y_j \mid \text{NOHVYCOLL}(H) \text{ and } \theta(i) \neq \theta(j)] = |f_i||f_j| \ .$$

Otherwise, let $t = \theta(i) = \theta(j)$. Then,

$$Y_i Y_j = \big(T_t[h_t(i)] \cdot \text{sgn}(f_i) \cdot \xi(i)\big) \cdot \big(T_t[h_t(j)] \cdot \text{sgn}(f_j) \cdot \xi(j)\big)$$
$$= \text{sgn}(f_i f_j) \sum_{\substack{i':h_t(i')=h_t(i) \\ j':h_t(j')=h_t(j)}} f_j f_{j'} \xi(j)\xi(j')\xi(i)\xi(i')$$

Note that since NOHVYCOLL(H) holds, $h_t(i) \neq h_t(k)$ and therefore, $i' \neq k'$. Taking expectations and using four-wise independence of the $\xi$'s obtain

$$\mathsf{E}[Y_i Y_k \mid \text{NOHVYCOLL}(H) \text{ and } \theta(i) = \theta(j)] = |f_i||f_j| \ .$$

Th Therefore, in all cases, we have

$$\mathsf{E}[Y_i Y_k \mid \text{NOHVYCOLL}(H)] = |f_i||f_j| \qquad\qquad i \neq j, i, j \in H \ . \quad \blacksquare \qquad (9)$$

**Lemma 3.8** *If $\epsilon \leq \frac{1}{4}$, $B = 1/\epsilon^2$, $C = 64B$ and $g = \log \frac{36B^2}{\epsilon^4}$, then $\Pr\left[|\hat{F}_1^H - F_1^H| \leq \epsilon F_1\right] \geq \frac{2}{3}$.*

**Proof** Let NoHvyColl be an abbreviation for the event NoHvyColl($H$). Let $|H| = m'$.

$$\mathsf{Var}_\xi\left[\sum_{i \in H} Y_i \mid \text{NoHvyColl}\right]$$
$$= \sum_{i \in H} \left(\mathsf{E}_\xi\left[Y_i^2 \mid \text{NoHvyColl}\right] - (\mathsf{E}_\xi\left[Y_i \mid \text{NoHvyColl}\right])^2\right)$$
$$+ \sum_{i,j \in H, i \neq j} \left(\mathsf{E}_\xi\left[Y_i Y_j \mid \text{NoHvyColl}\right] - \mathsf{E}_\xi\left[Y_i \mid \text{NoHvyColl}\right]\mathsf{E}_\xi\left[Y_j \mid \text{NoHvyColl}\right]\right)$$
$$= \sum_{i \in H} \sum_{\substack{k: h_{\theta(i)}(k) = h_{\theta(i)}(i) \\ k \neq i, k \notin H}} f_k^2 + 0, \quad \text{(by Lemma 3.7)} \quad .$$

Therefore $\mathsf{Var}_{h,\xi}\left[\sum_{i \in H} Y_i \mid \text{NoHvyColl}\right] = \frac{|H|}{C} F_2^{\text{res}}(H)$. As in [3], define the event

$$\text{LowVar} \equiv \mathsf{Var}_\xi\left[\sum_{i \in H} Y_i \mid \text{NoHvyColl}(H)\right] \leq \frac{8|H|F_2^{\text{res}}(H)}{C} \quad .$$

By Markov's inequality, $\mathsf{Pr}_h[\text{LowVar} \mid \text{NoHvyColl}] \geq \frac{7}{8}$. By Chebychev's inequality,

$$\mathsf{Pr}\left[\left|\sum_{i \in H} Y_i - \sum_{i \in H} |f_i|\right| \leq 8\left(\frac{|H|F_2^{\text{res}}(8B)}{C}\right)^{1/2} \mid \text{NoHvyColl and LowVar}\right] \geq \frac{7}{8} \quad .$$

Unconditioning dependencies,

$$\mathsf{Pr}\left[\left|\sum_{i \in H} Y_i - \sum_{i \in H} |f_i|\right| \leq 8\left(\frac{|H|F_2^{\text{res}}(B)}{16B}\right)^{1/2}\right] \geq \frac{7}{8}\mathsf{Pr}[\text{LowVar} \mid \text{NoHvyColl}]\mathsf{Pr}[\text{NoHvyColl}]$$
$$\geq \frac{7}{8} \cdot \frac{7}{8} \cdot \frac{31}{32} \geq \frac{2}{3} \quad . \tag{10}$$

Recall that $|H| \leq 5.1B$ and by Lemma 3.5, $F_2^{\text{res}}(H) \leq 12.04 F_2^{\text{res}}(|H|)$. Therefore,

$$\left(\frac{|H|F_2^{\text{res}}(H)}{64B}\right)^{1/2} \leq \left(\frac{12.04|H|F_2^{\text{res}}(|H|)}{64B}\right)^{1/2} \leq \left(\frac{12.04|H|}{64B|H|}\right)^{1/2} F_1 \leq \frac{0.44}{\sqrt{B}} F_1 = 0.44\epsilon F_1$$

Substituting in (10), we have $\mathsf{Pr}\left[\left|\sum_{i \in H} Y_i - \sum_{i \in H} |f_i|\right| \leq 3.6\epsilon F_1\right] \geq \frac{2}{3}$. ∎

### 3.3 Total Error

In this section, we add the various errors to obtain the total error of the estimate.

**Theorem 3.9** $|\hat{F}_1 - F_1| \leq 10\epsilon F_1$ *with prob. 0.576.*

**Proof** From analysis of light estimator (Lemma 3.4 and setting $p = 1$) we have

$$\left|\hat{F}_1^L - F_1^L\right| \leq 6\epsilon F_1 \text{ with probability } 35/36.$$

By heavy estimator (Lemma 3.8) we have

$$\left|\hat{F}_p^H - F_p^H\right| \leq 3.6\epsilon F_1 \text{ with prob. } \frac{2}{3}.$$

Since, $\hat{F}_1 = \hat{F}_1^H + \hat{F}_1^L$ and $F_1 = F_1^L + F_1^H$, we have

$$\left|\hat{F}_1 - F_1\right| \leq 10\epsilon F_1$$

with prob. $1 - \frac{2}{32} - \frac{1}{36} - \frac{1}{3} = 0.576$. ∎

## 3.4 Reducing Random Bits

We now reduce the randomness requirements for the stable sketches and the hash functions.

*Stable Sketches.* Using Nisan's PRG, a *single* stable sketch used in a bucket of a table $U$ may be fooled using Nisan's PRG using a seed length of $T = O\big((\log \frac{mM}{\epsilon})(\log n)\big)$ bits. The three stable sketches in each bucket need to be only 3-wise independent. The stable sketches used across the buckets of a table $U$ need to be only pair-wise independent to facilitate variance calculations. Thus, it suffices to use a pair-wise independent hash function $g$ that maps $3T$-bit strings to $3T$-bits strings. The seeds for each of the buckets is obtained as $g(1), g(2), \ldots, g(C)$. Each of the $3L$-bit string is viewed as the seed for 3-wise independent hash function $h'_b$. The number of random bits used per table is $3L$. The seeds for stable sketches across the tables are pair-wise independent, since the random variables are used only in variance calculations. Hence we can use a random seed length of $O(T) = O\big((\log \frac{mM}{\epsilon})(\log n)\big)$.

*Independence of hash functions.* There are two occasions where full independence properties of hash functions are used, namely, (i) for $i \in H$, $1/C_L = \Pr[\text{NoCollsion}(i)]$ is estimated as $(1 - 1/C)^{|H|-1}$, and, (ii) $\Pr[\text{NoHvyColl}(H)] \geq \frac{31}{32}$. Let $\Pr[\cdot]$ denote the probability of an event under full-independence of $h$ and let $\Pr_t[\cdot]$ denote the probability assuming the hash function is $t$-wise independent. Let $C_L^t$ denote $1/\Pr_t[\text{NoCollsion}(i)]$.

**Lemma 3.10** *If $t = \log \frac{1}{\epsilon^2}$, then, for any $i \in H$*

$$\left|C_L^{-1} - (C_L^t)^{-1}\right| = \left|\Pr_t[\text{NoCollsion}(i)] - \Pr[\text{NoCollsion}(i)]\right| \leq \epsilon^2 \ .$$

*If $t \geq 8$ and $g \geq 3 + \log(\epsilon^{-2})$ then, $\Pr_t[\text{NoHvyColl}] \geq 31/32$.*

**Proof** See Appendix A.

## 3.5 Space and Update time

In this section we summarize the resource consumption of the algorithm.

The space requirement is $O\big(\epsilon^{-2}(\log n + (\log \frac{1}{\epsilon}))\log(mM)\big)$. The length of the random seed is $O\big((\log \frac{mM}{\epsilon})(\log n) + \log \frac{1}{\epsilon}\big)$ and does not dominate the space requirement. The time taken to update the $HH_2$ structure is $O(\log n)$. The hash tables $T_j$ and $U$ use 8-wise independent hash functions (except $T_1$ and $U_1$ that use $O(\log \frac{1}{\epsilon})$-wise independence for estimating $F_1^L$). The hash values are calculated in time $O(g) = O(\log \frac{1}{\epsilon})$. Nisan's PRG is used to generate a chunk of size $3T$ bits using $O((\log n)(\log \frac{1}{\epsilon}))$ operations on word size $O\big(\log(mM)\big)$ bits. The total update time is $O\big((\log n)(\log \frac{1}{\epsilon}) + (\log \frac{1}{\epsilon}) + (\log n)\big) = O((\log n)(\log \epsilon^{-1}))$.

## 4 Conclusion

We first present a novel space-optimal algorithm for estimating $F_p$ over data streams to within multiplicative error factor of $1 \pm \epsilon$ for $p \in (0, 2]$. We then present an algorithm for estimating $F_1$. This algorithm is nearly optimal with respect to both space usage and update processing time. The space requirement of the algorithm is $O(\epsilon^{-2} \log(n\epsilon^{-1}) \log(mM))$ and a per-update processing time of $O((\log n)(\log \epsilon^{-1}))$.

## A  Proofs

**Proof of Claim 1** $\hat{f}_i \in f_i \pm \left(\frac{F_2^{\text{res}}(8B)}{B}\right)^{1/2} = \Delta$ (say). For $i \in H$,

$$f_i \leq \hat{f}_i + \Delta \leq \left(\frac{\hat{F}_2^{\text{res}}(\epsilon B)}{\epsilon B}\right)^{1/2} + \left(\frac{F_2^{\text{res}}(B)}{B}\right)^{1/2} \leq \left(\frac{F_2^{\text{res}}(\epsilon B)}{\epsilon B}\right)^{1/2}(\sqrt{33/32} + \sqrt{\epsilon}), \quad i \in H, \text{ and}$$

$$f_i \geq 2\left(\frac{31 F_2^{\text{res}}(4B)}{32B}\right)^{1/2} - \Delta \geq 2\left(\frac{31 F_2^{\text{res}}(4B)}{32B}\right)^{1/2} - \left(\frac{F_2^{\text{res}}(8B)}{B}\right) \geq 1.04\left(\frac{F_2^{\text{res}}(4B)}{B}\right)^{1/2}, \quad i \in H \ .$$

Therefore,

$$|H| \leq 4B + (1.04)^2 B \leq 5.1 B \ . \qquad \blacksquare \tag{11}$$

**Proof of Lemma 3.1** Divide the items in order of consecutive groups $G_1, G_2, \ldots, G_{\lceil n/B \rceil}$ of size $B$ items each, that is, $G_1$ contains the first $B$ items in non-increasing order of absolute frequency values, $G_2$ contains the next $B$ items, and so on. The last group may contain fewer than $B$ items. Let $q \geq p$.

$$\sum_{j=B+1}^{n} |f_{s_i}|^q = \sum_{l=2}^{\lceil n/B \rceil} \sum_{i \in G_l} |f_{s_i}|^q$$

$$\leq \sum_{l=2}^{\lceil n/B \rceil} \sum_{i \in G_l} \left(\frac{1}{B} \sum_{i \in G_{l-1}} |f_{s_i}|^p\right)^{q/p}, \quad \text{for } i \in G_l, |f_{s_i}|^p \leq \text{avg}\{|f_j|^p : j \in G_{l-1}\}, p \geq 0$$

$$\leq \sum_{l=1}^{\lceil n/B \rceil - 1} \frac{1}{B^{q/p-1}} \left(\sum_{i \in G_l} |f_{s_i}|^p\right)^{q/p} \leq \frac{1}{B^{q/p-1}} \left(\sum_{j=1}^{n} |f_{s_i}|^p\right)^{q/p}, \quad \text{since, } q \geq p \ .$$

The particular case is obtained by setting $q = 2p$ in the above equation. $\blacksquare$

**Proof of Lemma 3.10.** Fix a table index $j \in [g]$ and an item $i \in H$. Let $k \in H$, $k \neq i$. Define the indicator variable $x_k$ to be 1 if $k$ collides with $i$ in the same bucket in table $U$, that is, $h_j(i) = h_j(k)$. Let $Y = \sum_{k \in H, k \neq i} x_k$. The event $\text{NoCollsion}(i)$ is equivalent to $Y = 0$. Let $\mu = \mathsf{E}[Y] = \frac{|H|-1}{C} \leq \frac{5.1 B}{64 B} \leq 0.1$.

By Theorem 2.6, part (III) of [12] (proved using inclusion-exclusion), if $t \geq e\mu + \ln(1/\Pr[Y = 0]) + 1 + D$, then,

$$\left|\Pr_t[Y \geq 1] - \Pr[Y \geq 1]\right| \leq (1 - \Pr[Y \geq 0]) e^{-D} \ .$$

We have, $\Pr[Y = 0] = (1 - 1/C)^{|H|-1} \leq 2(|H|-1)/C \leq 1/5$. Therefore, for $t \geq 0.1 e + \ln(5) + 1 + D$

$$\left|\Pr_t[Y = 0] - \Pr_t[Y = 0]\right| = \left|\Pr_t[Y \geq 1] - \Pr[Y \geq 1]\right| \leq (4/5) e^{-D} \ .$$

It suffices for the *RHS* to be $\epsilon^2$, which can be satisfied by keeping $D = \log(1/\epsilon^2)$.

For $t \geq 8$,

$$\Pr_t[\text{NoHvyColl}] \geq 1 - |H|\left(1 - \left(1 - \Pr_t[\text{NoCollsion}(i)]\right)^g\right)$$

$$\geq 1 - |H|\left(1 - \left(1 - \Pr[\text{NoCollsion}(i)] - (4/5) e^{-6}\right)^g\right)$$

$$\geq \frac{31}{32}$$

provided, $g \geq \log(5.1 B) \geq 3 + \log(\epsilon^{-2})$. $\blacksquare$